\documentclass[12pt]{article}
\let\section=\subsection     \let\subsection=\subsubsection                
\usepackage{graphicx}

\usepackage{feynmp}

\usepackage{amssymb}
\usepackage{amsmath}

\usepackage{xspace}             
\newcommand{\ia}{{\"{\i}}}

\newcommand{\blue}[1]{#1}
\newcommand{\red}[1]{#1}

\newcommand{\green}[1]{#1}

\newcommand{\be}{\begin{equation}}
\newcommand{\ee}{\end{equation}}
\newcommand{\bea}{\begin{eqnarray}}
\newcommand{\eea}{\end{eqnarray}}

\newcommand{\half}{\frac{1}{2}}

\newcommand{\ii}{\mathrm{i}}

\newcommand{\fm}{\;\mathrm{fm}}
\newcommand{\MeV}{\;\mathrm{MeV}}

\newcommand{\de}{\partial}

\newcommand{\EPJA}{\textit{Eur.\ Phys.\ J.\ }\textbf{A}}

\newcommand{\PR}{\textit{Phys.\ Rev.\ }}

\newcommand{\PRC}{\PR\textbf{C}}

\fmfstraight

\newcommand{\hq}{\hspace*{0.5em}}

\newlength{\feyngraphlength}

\begin{document}
\begin{center}
  {\large \bf Compton Scattering off Nucleons:
  }\\[2mm]
  {\large \bf Focus on 
    the Nucleonic Low-Energy Degrees of Freedom}\\[5mm]
  Harald W.~Grie\3hammer\footnote{Email: hgrie@physik.tu-muenchen.de.
    Supported by DFG under GR 1887/2-1.}~\footnote{Preprint TUM-T39-04-03,
    nucl-th/0402052. Talk given at the \textsc{XXXIIth International Workshop
      on Gross Properties of Nuclei and Nuclear Excitations: Probing Nuclei
      and Nucleons with Electrons and Photons},
    Hirschegg (Austria), 11th -- 17th January 2004. To be published in the
    proceedings.}\\[5mm] 
  {\small \it T39, Physik-Department, TU M{\"u}nchen, D-85747 Garching,
    Germany\\[8mm]
  }
\end{center}

\begin{abstract}\noindent
  In this sketch, I focus on Physics and formalism behind dynamical
  polarisabilities, a new tool to test and interpret quantitative predictions
  about the low-energy degrees of freedom inside the nucleon from the
  multipoles of Compton scattering for photon energies below $300$ MeV. A way
  to extract them from double-polarised precision experiments is sketched.
  Predictions from Chiral Effective Field Theory, both on the proton and on
  the neutron, serve as guideline for forthcoming experiments. Special
  interest is put on the r\^ole of the nucleon spin-polarisabi\-lities. For
  details and a complete list of references, consult
  Refs.~\cite{Griesshammer:2001uw,polas2,polas3}.
\end{abstract}

\section{Introduction}
Nuclear physicists are hardly surprised by the fact that in low-energy Compton
scattering $\gamma N\to\gamma N$, the nucleon is not a point-like spin-$\half$
target with some anomalous magnetic moment. In fact, these nucleon structure
effects have been known for many decades and (in the case of a proton target)
quite reliable theoretical calculations for the deviations from the
Powell-cross section exist. They are canonically parameterised starting from
the most general interaction between the nucleon and an electro-magnetic field
of fixed, non-zero energy $\omega$:
\begin{equation}
\begin{split}
  &2\pi\;\Bigg\{\bigg[\red{\alpha_{E1}(\omega)}\;\vec{E}^2\;+
  \;\red{\beta_{M1}(\omega)}\;\vec{B}^2\bigg]\;+
  \;\frac{1}{6}\;\bigg[\red{\alpha_{E2}(\omega)}\;E_{ij}^2\;+
  \;\beta_{M2}(\omega)\;B_{ij}^2\bigg]
  \\&\;\;\;\;\;\;\;\;
  \label{polsfromints}
  +\;\bigg[\red{\gamma_{E1E1}(\omega)}
  \;\vec{\sigma}\cdot(\vec{E}\times\dot{\vec{E}})\;
  +\;\red{\gamma_{M1M1}(\omega)}
  \;\vec{\sigma}\cdot(\vec{B}\times\dot{\vec{B}}) \\&\;\;\;\;\;\;\;\;\;\;\;\;
  -\;2\red{\gamma_{M1E2}(\omega)}\;\sigma_i\;B_j\;E_{ij}\;+
  \;2\red{\gamma_{E1M2}(\omega)}\;\sigma_i\;E_j\;B_{ij} \bigg] \;+\;\dots
   \Bigg\}\;N^\dagger N
\end{split}
\end{equation}
Here, the electric or magnetic ($\blue{X,Y=E,M}$) photon undergoes a
transition $\blue{Xl\to Yl^\prime}$ of definite multipolarity
$\blue{l,l^\prime=l\pm\{0,1\}}$; $\green{T_{ij}:=\half (\de_iT_j +
  \de_jT_i)}$. Thus, there are six dipole polarisabilities: two
spin-independent ones ($\alpha_{E1}(\omega)$, $\beta_{M1}(\omega)$) for
electric and magnetic dipole transitions which do not couple to the nucleon
spin; and in the spin sector, two diagonal (``pure'') spin-polarisabilities
($\gamma_{E1E1}(\omega)$, $\gamma_{M1M1}(\omega)$), and two off-diagonal
(``mixed'') spin-polarisabilities, $\gamma_{E1M2}(\omega)$ and
$\gamma_{M1E2}(\omega)$. In addition, there are higher ones like quadrupole
and octupole polarisabilities, with negligible contributions, see next
Section.

Each of these quantities parameterises the global stiffness of the nucleon's
internal degrees of freedom against displacement in an electric or magnetic
field of definite multipolarity and non-vanishing frequency $\omega$. They are
energy-dependent because different mechanisms (low-lying nuclear resonances
like the $\Delta(1232)$, the charged meson cloud around the nucleon etc.)
react quite differently to real photon fields of non-zero frequency.
Therefore, these \emph{dynamical polarisabilities} contain detailed
information about dispersive effects, caused by internal relaxation, baryonic
resonances and mesonic production thresholds.

Nucleon Compton scattering provides thus a wealth of information about the
internal structure of the nucleon. However, in contradistinction to many other
electro-magnetic processes -- e.g.~pion photo-production off a nucleon -- the
nucleon structure effects probed in Compton scattering in most of the recent
analyses have not been analysed in terms of a multipole expansion. Instead,
most experiments have focused on just two structure parameters, namely the
static electric and magnetic polarisabilities
$\bar{\alpha}_E:=\alpha_{E1}(\omega=0)$ and
$\bar{\beta}_M:=\beta_{M1}(\omega=0)$. Therefore, at present, quite different
theoretical frameworks are able to provide a consistent, qualitative picture
for the leading static polarisabilities. Their dynamical origin is however
only properly revealed by their energy-dependence, which varies from model to
model. Even less is known about the spin-polarisabilities, see Sect.~4.

\section{Definition of Dynamical Polarisabilities}

A rigorous definition of the energy-dependent or dynamical polarisabilities
starts instead of (\ref{polsfromints}) from the six independent amplitudes
into which the $T$-matrix of real Compton scattering is decomposed:
\begin{equation}
\begin{split}
  T(\omega,z)&= A_1(\omega,z)\,\vec{\epsilon}\,'^*\cdot \vec{\epsilon} +
  A_2(\omega,z)\,
  \vec{\epsilon}\,'^*\cdot\hat{\vec{k}}\;\vec{\epsilon}\cdot\hat{\vec{k}}'
  \\
  & +\ii\,A_3(\omega,z)\,\vec{\sigma}\cdot
  \left(\vec{\epsilon}\,'^*\times\vec{\epsilon}\,\right)
  +\ii\,A_4(\omega,z)\,\vec{\sigma}\cdot
  \left(\hat{\vec{k}}'\times\hat{\vec{k}}\right)
  \vec{\epsilon}\,'^*\cdot\vec{\epsilon}
  \\
  & +\ii\,A_5(\omega,z)\,\vec{\sigma}\cdot
  \left[\left(\vec{\epsilon}\,'^*\times\hat{\vec{k}} \right)\,
    \vec{\epsilon}\cdot\hat{\vec{k}}' -\left(\vec{\epsilon}
      \times\hat{\vec{k}}'\right)\,
    \vec{\epsilon}\,'^*\cdot\hat{\vec{k}}\right]\\
  &+\ii\,A_6(\omega,z)\,\vec{\sigma}\cdot
  \left[\left(\vec{\epsilon}\,'^*\times\hat{\vec{k}}'\right)\,
    \vec{\epsilon}\cdot\hat{\vec{k}}' -\left(\vec{\epsilon}
      \times\hat{\vec{k}} \right)\,
    \vec{\epsilon}\,'^*\cdot\hat{\vec{k}}\right]
\end{split} 
\label{Tmatrix}
\end{equation}
Here, $\hat{\vec{k}}$ ($\hat{\vec{k}}'$) is the unit vector in the momentum
direction of the incoming (outgoing) photon with polarisation $\vec{\epsilon}$
($\vec{\epsilon}\,'^*$).

We separate these amplitudes into a pole and non-pole or structure
($\bar{A}_i$) part. Intuitively, one could define the pole part as the one
which leads to the Powell cross section of a point-line nucleon with anomalous
magnetic moment and thus parameterises all we hope to have understood about
the nucleon. Then, it would seem, the structure part contains all information
about the internal degrees of freedom which make the nucleon an extended,
polarisable object. However, the question which part a contribution belongs to
cannot be answered uniquely. In the following, only those terms which have a
pole either in the $s$-, $u$- or $t$-channel are treated as non-structure.  In
the calculation of observables, this separation is clearly irrelevant because
both the structure dependent and structure independent part contribute. Here,
however, we investigate the r\^ole of the internal nucleonic degrees of
freedom on the polarisabilities, which are contained only in the structure
part of the amplitudes.

We also choose to work in the centre-of-mass frame. Thus, $\omega$ denotes the
cm energy of the photon, $M$ the nucleon mass, $W=\sqrt{s}$ the total
cm-energy, and $\theta$ the cm-scattering angle, $z=\cos\theta$. Following
older work on the multipole-decomposition of the Compton amplitudes and
pulling a kinematical factor out relative to (\ref{polsfromints}), one obtains
for the expansion of the \emph{structure parts} of the amplitudes in terms of
polarisabilities
\begin{align}
  \bar{A}_1(\omega,\,z)
  &\textstyle=\frac{4\pi\,W}{M}\,\left[\alpha_{E1}(\omega)
    +z\,\beta_ {M1}(\omega)\right]\,\omega^2+\dots
  , \nonumber
  \\[1ex]
  \bar{A}_2(\omega,\,z) &\textstyle
  =-\frac{4\pi\,W}{M}\,\beta_{M1}(\omega)\,\omega^2
  +\dots
  , \nonumber
  \\[1ex]
  \bar{A}_3(\omega,\,z)
  &\textstyle=-\frac{4\pi\,W}{M}\,\left[\gamma_{E1E1}(\omega)
    +z\,\gamma_{M1M1}(\omega)+\gamma_{E1M2}(\omega)
    +z\,\gamma_{M1E2}(\omega)\right]\,\omega^3+\dots
  , \nonumber
  \\[1ex]
  \bar{A}_4(\omega,\,z) &\textstyle=\frac{4\pi\,
    W}{M}\,\left[-\gamma_{M1M1}(\omega)
    +\gamma_{M1E2}(\omega)\right]\,\omega^3+\dots
  ,
\label{eq:strucamp}
\\[1ex]
\bar{A}_5(\omega,\,z) &\textstyle=\frac{4\pi\,
  W}{M}\,\gamma_{M1M1}(\omega)\,\omega^3
  +\dots
  , \nonumber
  \;\;\;\; \bar{A}_6(\omega,\,z) =\frac{4\pi\,
    W}{M}\,\gamma_{E1M2}(\omega)\,\omega^3+\dots
\end{align}
The various polarisabilities are thus identified \emph{at fixed energy} only
by their different angular dependence. Clearly, the complete set of dynamical
polarisabilities does not contain more or less information about the temporal
response or dispersive effects of the nucleonic degrees of freedom than the
un-truncated amplitudes. However, the information is more readily accessible:
We will be able to see directly which Physics can be found in which
polarisability. Moreover, it will turn out that all polarisabilities beyond
the dipole ones can be dropped in (\ref{eq:strucamp}), as they are so far
invisible in observables. This is why they were sacrificed to brevity in the
expressions above. Purists consider Ref.~\cite{polas2}.

\section{Low-Energy Contents of Dynamical Polarisabilities}

Which energy-dependent effects can we expect? Polarising the pion cloud around
the nucleon should result in a characteristic cusp as one approaches the
one-pion production threshold. It is also well-known that the $\Delta(1232)$
as the lowest nuclear resonance leads to a large para-magnetic contribution to
the static magnetic dipole polarisability~\footnote{As is common practise, we
  measure the scalar dipole polarisabilities in units of $10^{-4}\;\fm^3$}
$\bar{\beta}_{M1}^{\Delta}=+[7\dots13]$ due to its strong $M1\to M1$
transition. A characteristic resonance shape should occur, like predicted by
the Lorentz model of polarisabilities in classical electro-dynamics.  As the
observed static value $\bar{\beta}_{M1}\approx 1.5$ is smaller by a factor of
$5$ to $10$ than the $\Delta$ contribution, some other mechanism must provide
a strong dia-magnetic component. The resultant fine-tuning at zero photon
energy is unlikely to hold once we consider the evolution of the
polarisabilities as a function of the photon energy: If dia- and
para-magnetism are of different origin, it is more than likely that they
involve different scales and hence different energy-dependences. Therefore,
they are apt to be dis-entangled when one extends static polarisabilities to
the non-zero energy range, i.e.~to the \emph{dynamical polarisabilities}.

\begin{center}
  \includegraphics*[width=\linewidth]{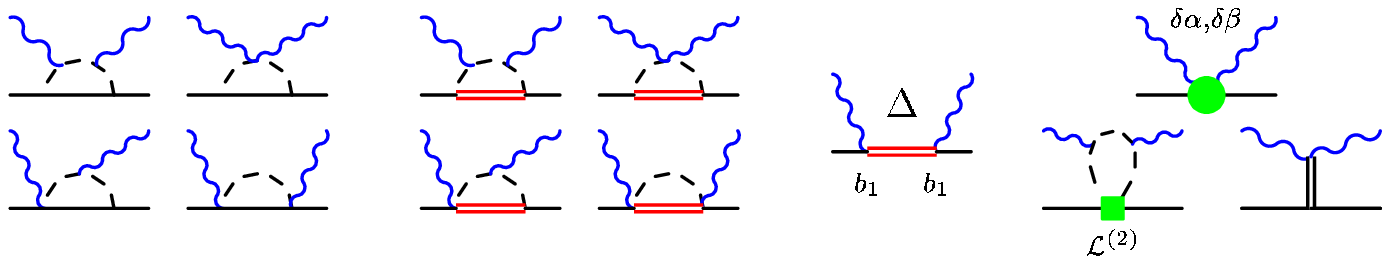}
  
  \vspace*{-1ex}
            
  \parbox{14cm}{\footnotesize Fig.~1: Dominant interactions in $\chi$EFT which
    give rise to the nucleon polarisabilities, left to right: pion cloud
    around the nucleon and $\Delta$; $\Delta$ excitations; short-distance
    effects (upper right) with two possible explanations (lower right):
    higher-order $\chi$EFT~\cite{BKMS}, or meson-exchange via the
    $t$-channel~\cite{Guichon}. Permutations and crossed diagrams not shown.
    From Ref.~\cite{polas2}.}
\end{center}

To identify the microscopically dominant low-energy degrees of freedom inside
the nucleon in a model-independent way, we employ the Chiral Effective Field
Theory ($\chi$EFT) of QCD in the one-nucleon system. This extension of Chiral
Perturbation Theory to the few-nucleon sector contains only the observable
low-energy degrees of freedom, interacting in all ways allowed by the
underlying symmetries of QCD. A power counting allows for results of finite,
systematically improvable accuracy, and thus an error estimate. The
contributions at leading-one-loop order are, see Fig.~1: photons coupling to
the pion cloud around the nucleon and the $\Delta$, and excitation of the
$\Delta$ as intermediate state by the $\gamma N\Delta$-coupling $b_1$.
Finally, all short-distance Physics not generated by these degrees of freedom
is sub-sumed into two low-energy coefficients
$\delta\alpha_{E1},\;\delta\beta_{M1}$, which are \emph{energy-independent}.
$\chi$EFT also predicts that the proton and neutron polarisabilities are very
similar, iso-vectorial effects being higher order in the power counting.

\begin{center}\footnotesize
\begin{tabular}{|c||r|r|r|}
\hline
Quantity&3-parameter-fit&2-parameter-fit&Olmos 2001\\
\hline
$\chi^2/d.o.f.$ &2.87&
                 2.83&1.14 \\
\hline
$\bar{\alpha}_E\;\,[10^{-4}\,\mathrm{fm}^3]$&$11.52\pm2.43$
                 &$11.04\pm1.36$
                 &$12.4\pm0.6(\mathrm{stat})\mp0.6(\mathrm{sys})$\\
$\bar{\beta} _M\;\,[10^{-4}\,\mathrm{fm}^3]$&$ 3.42\pm1.70$
                 &$2.76\mp1.36$
                 & $1.4\pm0.7(\mathrm{stat})\pm0.5(\mathrm{syst})$\\
$b_1              $&$4.66\pm0.14$ 
                   &$4.67\pm0.14$&\\
\hline
\end{tabular}\\[2ex]
\parbox{14cm}{\footnotesize Table 1: The static polarisabilities
  $\bar{\alpha}_E$, $\bar{\beta}_M$ and $b_1$ from a fit to MAMI- and
  SAL-data, compared to the results from Mainz. From Ref.~\cite{polas2}. }
\end{center}

The three constants $b_1$, $\delta\alpha_{E1}$ and $\delta\beta_{M1}$ are
determined by fitting the un-expanded, complete $\chi$EFT-amplitude to the
cornucopia of Compton scattering data on the proton, cf.~Fig.~2. Table~1 shows
that the results are in good agreement with state-of-the-art results from
Dispersion Theory~\cite{Report}, with comparable error bars. The value of
$\bar{\alpha}_E+\bar{\beta}_M$ from the three-parameter-fit is consistent
within error bars with the Baldin sum rule for the proton,
$\bar{\alpha}_E+\bar{\beta}_M=13.8$. One can therefore in a second step use
the value of the Baldin sum rule as additional data point and reduce the
number of free parameters to two, as done in the following. The value for
$b_1$ is also consistent with the one obtained from the radiative
$\Delta$-decay-width. Albeit a na{\ia}ve dimensional estimate predicts the two
short-distance parameters to be small in magnitude, $\lesssim 1.5$, they are
anomalously large, $\delta\alpha_{E1}=-5.9\pm1.4,\;
\delta\beta_{M1}=-10.7\pm1.2$, justifying their inclusion at leading order. As
expected, $\delta\beta_{M1}$ is dia-magnetic.

\begin{center}
  \includegraphics[width=0.4\linewidth,bb=111 495 285
  611,clip=]{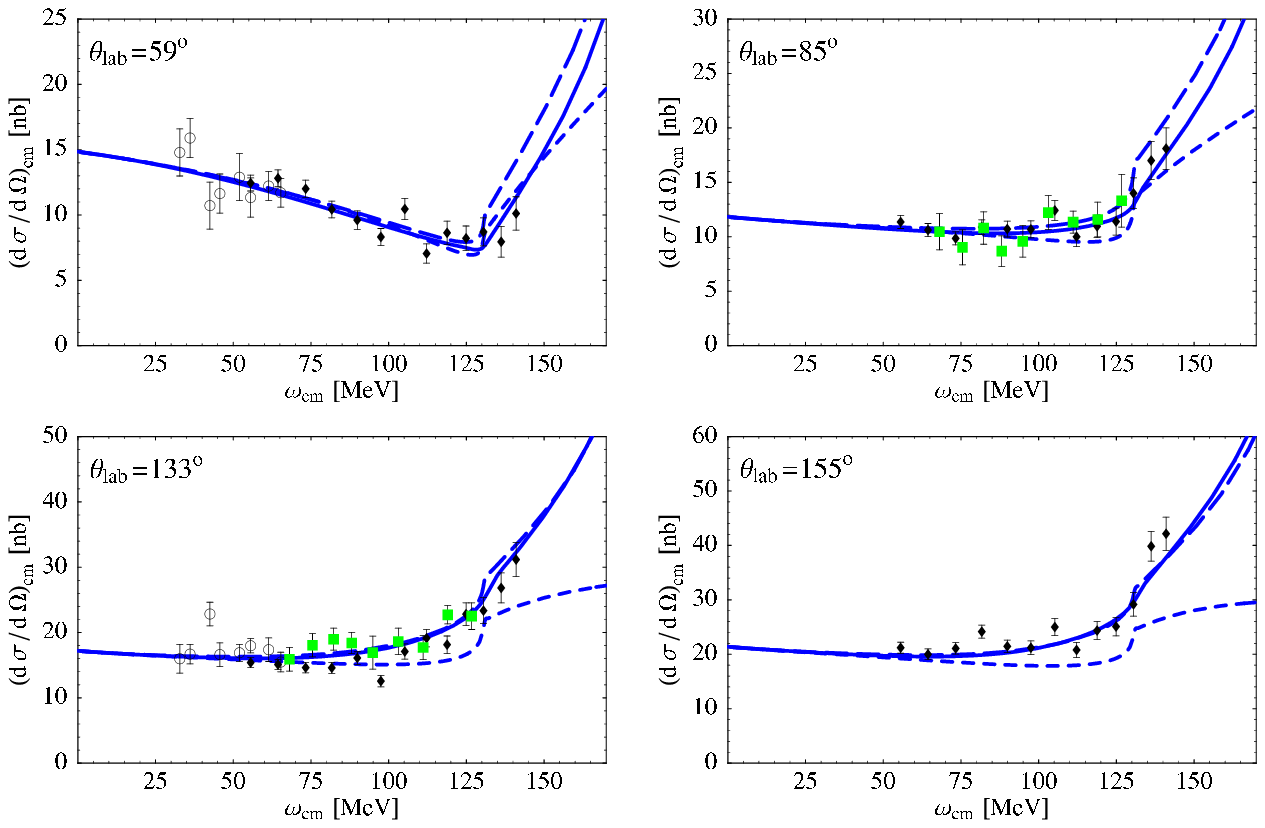} \hq\hq
  \includegraphics[width=0.4\linewidth,bb=111 495 285
  611,clip=]{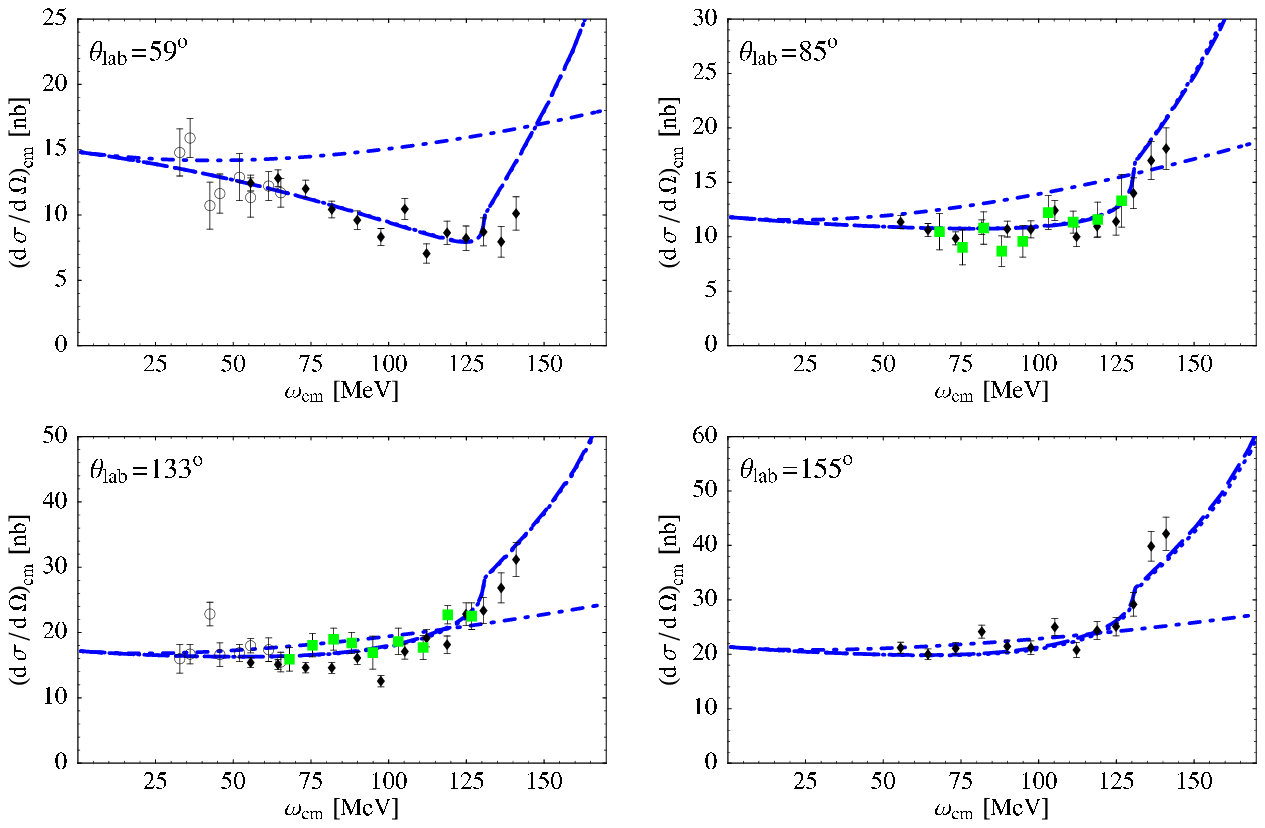}\\[1ex]
  \parbox{14cm} {\footnotesize Fig.~2: Left: Typical differential cross
    section data for Compton scattering off the proton compared to Dispersion
    Theory (solid line) and $\chi$EFT with (long dashed) and without (short
    dashed) $\Delta$ contributions. Right: Data and $\chi$EFT without
    polarisabilities (dash-dotted), with only dipole polarisabilities
    (dashed), with full amplitude (dotted). From Ref.~\cite{polas2}.}
\end{center}

As the influence of the quadrupole and higher polarisabilities on cross
sections and asymmetries for energies up to about $300\MeV$ is hardly visible,
cf.~Figs.~2 and 4, truncating the multipole expansion in (\ref{eq:strucamp})
is justified.

With the parameters now fixed, the energy-dependence of all polarisabilities
is predicted.  We compare with a result from dispersion theory, in which the
energy-dependent effects are sub-sumed into integrals over experimental input
from a different kinematical r\'egime, namely photo-absorption cross-section
$\gamma N\to X$. Its major source of error is the insufficient neutron data,
and the uncertainty in modelling the high-energy behaviour of the dispersive
integral.

The dipole polarisabilities, Fig.~3, show the behaviour expected above.  The
strong energy-dependence induced by the $\Delta$-resonance into the
polarisabilities containing an $M1$ photon reveals the good quantitative
agreement between the measured value of $\bar{\beta}_M$ and the prediction in
a $\chi$EFT without explicit $\Delta$ as accidental. For the first time, one
sees the unique $\Delta$ signature of a resonance-shape in
$\beta_{M1}(\omega)$ even well below the pion production threshold. While the
fine details of the rising para-magnetism differ between $\chi$EFT and
Dispersion Theory, they are consistent within the uncertainties of the
$\chi$EFT curve. The discrepancy between the two schemes above the one-pion
production threshold is likely to be connected to a detailed treatment of the
width of the $\Delta$-resonance, which is neglected in leading-one-loop
$\chi$EFT.  The pion-cusp -- so pronounced in the $E1$-polarisabilities -- is
quantitatively reproduced at leading order already. The spin-polarisabilities
are predictions, three of them being completely independent of the
parameter-determination. No genuinely new low-energy degrees of freedom inside
the nucleon are missing. Since the mixed spin-polarisabilities (lower panel of
Fig.~3) are small, the uncertainties of both Dispersion Theory and $\chi$EFT
are large there. More on that in the next Section.

\begin{center}
  \parbox[b]{0.31\linewidth}{ \includegraphics[width=\linewidth,bb= 111 614
    283
    731,clip=]{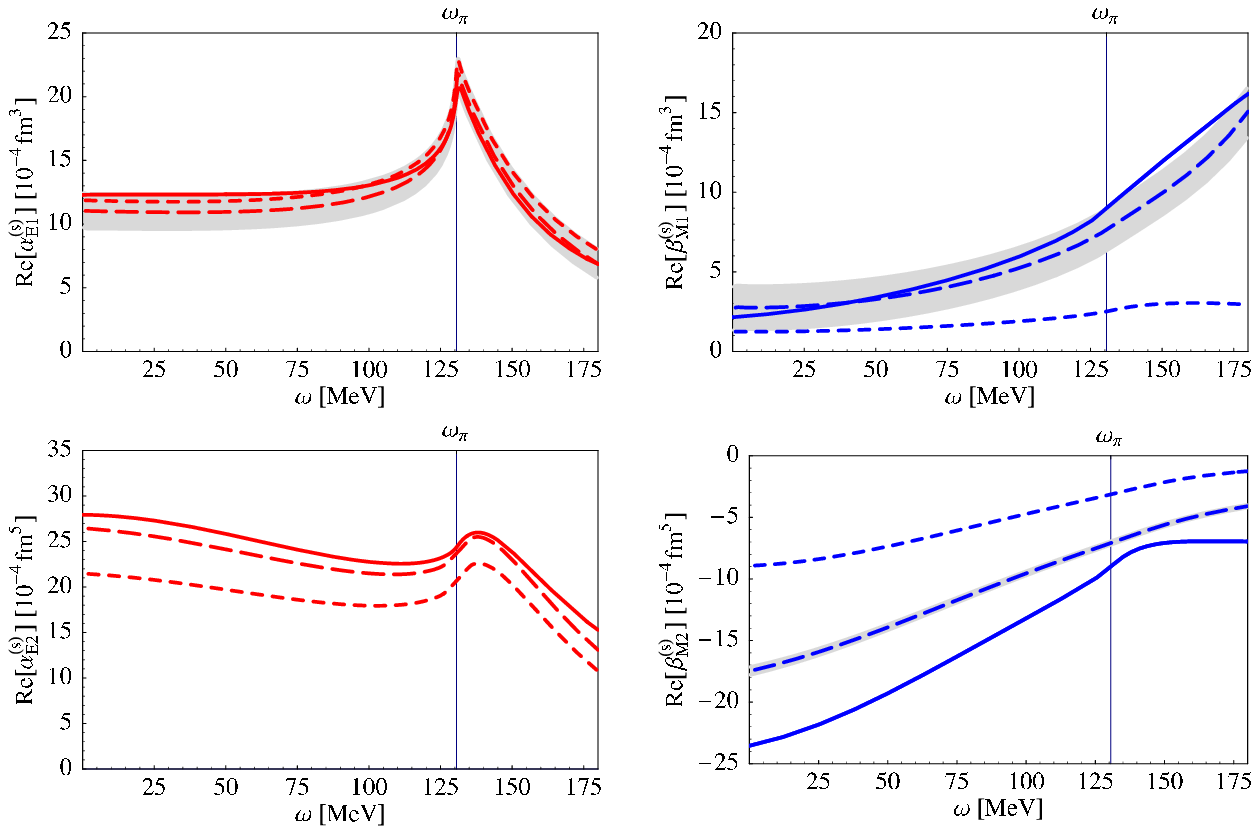}\\
    \includegraphics[width=\linewidth,bb= 298 614 470
    731,clip=]{spinindiepolas_isos_comparison_b1fit.eps} } \hfill
  \includegraphics[width=0.65\linewidth]{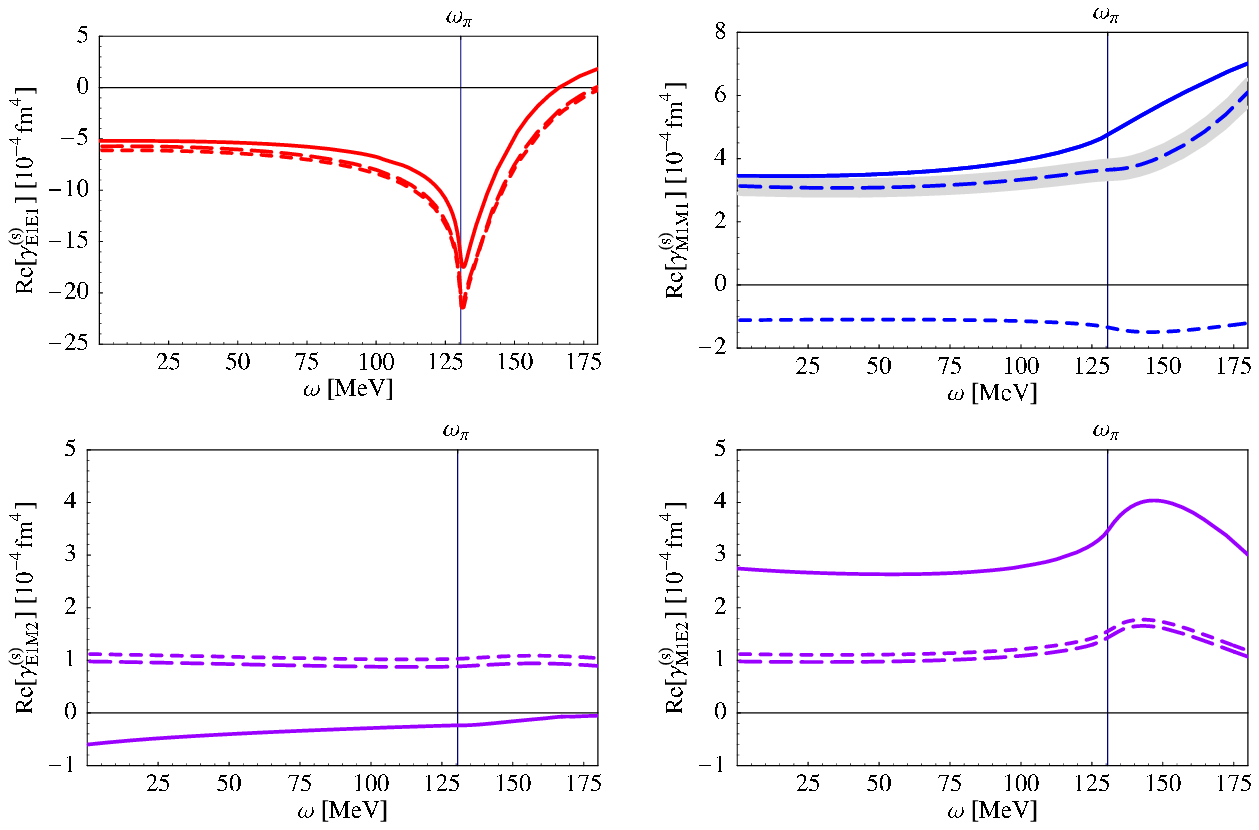}
  \\[1ex]
  \parbox{14cm} {\footnotesize Fig.~3: Dipole polarisabilities, predicted by
    Dispersion Theory (solid) and $\chi$EFT with (long dashed $+$ band from
    fit errors) and without (short dashed) explicit $\Delta$. Left:
    spin-independent; middle and right: spin-dependent. $\omega_\pi$ denotes
    the one-pion production threshold. From Ref.~\cite{polas2}.}
\end{center}

While the static polarisabilities of the nucleon are real, the dynamical
polarisabilities become complex once the energy in the intermediate state is
high enough to create an on-shell intermediate state, the first being the
physical $\pi N$ intermediate state, see~\cite{polas2}.

The two short-distance parameters $\delta\alpha_{E1},\;\delta\beta_{M1}$ which
sub-sume all Physics not generated by the pion cloud or the $\Delta$ suffice
to describe the polarisabilities up to energies of $300\MeV$ when the finite
width of the $\Delta$ is included. Therefore, three constraints arise on any
attempt to explain them microscopically:

\vspace*{-1.5ex}

\begin{itemize}
  \setlength{\itemsep}{0em} \setlength{\topsep}{0em} \setlength{\parsep}{0em}
\item[(1)] The effect must be $\omega$-independent over a wide range, like
  $\delta\alpha_{E1},\;\delta\beta_{M1}$.
  
  \vspace*{-0.5ex}
  
\item[(2)] It must occur in the electric and magnetic scalar polarisabilities,
  leading to the values for $\delta\alpha_{E1},\;\delta\beta_{M1}$ predicted
  in $\chi$EFT, but it must be absent in the pure spin-polarisabilities
  $\gamma_{E1E1},\;\gamma_{M1M1}$.
  
  \vspace*{-0.5ex}
  
\item[(3)] Its prediction for the proton and neutron must be very similar,
  because iso-vectorial effects were shown to be small and
  energy-independent~\cite{polas2}.
\end{itemize}

\vspace*{-1.5ex}

\noindent
Two proposals to explain $\delta\alpha_{E1},\;\delta\beta_{M1}$ were put
forward, see right side of Fig.~1. One attributes them to an interplay between
short-distance Physics and the pion cloud occuring from the next-to-leading
order chiral Lagrangean~\cite{BKMS}; the other to the $t$-channel exchange of
a meson or correlated two-pion exchange~\cite{Guichon}. Whether either of
these gives a convincing quantitative description of the short-distance
coefficients is not clear yet.

\section{Energy-Dependent Polarisabilities from Experiment}

Most experiments to determine polarisabilities are performed by Compton
scattering off protons and light nuclei at photon energies of $80-
200\;\mathrm{MeV}$. Fig.~3 shows that there, dynamical effects are large and
one can not just Taylor-expand the polarisabilities around their
zero-photon-energy value. Especially at large backward angle, unpolarised and
polarised cross-sections are rather sensitive to the non-analytical structure
of the amplitude around the pion cusp and $\Delta$-resonance, see Figs.~2 and
4, and~\cite{polas2,polas3}. The dipole spin-polarisabilities are anything but
negligible, even in un-polarised experiments.

While our knowledge about the (static) spin-independent polarisabilities is
rich, little information exists at present on the nucleonic
spin-polarisa\-bilities, which parameterise the response of the nucleon
\emph{spin} and its dominant low-energy degrees of freedom on an external
electro-magnetic field. Only two linear combinations are constrained from
experiments~\cite{Report}, and only at zero photon energy, namely the forward
and backward spin-polarisabilities $\bar{\gamma}_0$ and $\bar{\gamma}_\pi$ of
the nucleon which however involve all four static (dipole)
spin-polarisabilities.

As quadrupole and higher polarisabilities are negligible, one can use the
multipole-expansion of the scattering amplitudes in (\ref{eq:strucamp}) to
perform with increasing sophistication fits of the six dipole polarisabilities
per nucleon to data-sets which combine polarised and spin-averaged
experiments, taken at fixed energy but varying scattering angle.  One can for
example assume that the energy-dependence of the polarisabilities derived in
$\chi$EFT is correct: At low energies, only $\Delta(1232)$ and pion degrees of
freedom are expected to give dispersive contributions to the polarisabilities.
As starting values for the fit, one might thus use the
$\chi$EFT-results~\cite{polas2}, with deviations taken as energy-independent,
corresponding to a free normalisation for each dipole polarisability. Thus,
one obtains the dipole polarisabilities at a definite energy. Repeating this
procedure for various energies gives their energy dependence~\cite{polas3}.
This is one way to extract dynamical polarisabilities directly from the
angular dependence of observables.

As Fig.~3 shows, the results for the spin-independent polarisabilities
$\alpha_{E1}(\omega)$, $\beta_{M1}(\omega)$ from $\chi$EFT and Dispersion
Theory agree very well with each other, both in their energy-dependence and
overall size. They could therefore be used in a second step as input to reduce
the number of fit functions in (\ref{eq:strucamp}) to four, namely the four
dipole spin-polarisabilities. The good agreement in $\gamma_{E1E1}(\omega)$
and maybe even $\gamma_{M1M1}(\omega)$ can -- similarly -- be used
to reduce the number of fit functions further to three or two per nucleon:
$\gamma_{E1M2}(\omega)$ and $\gamma_{M1E2}(\omega)$.

\begin{center}
  \includegraphics[width=0.4\linewidth,bb=298 615 472
  730,clip=]{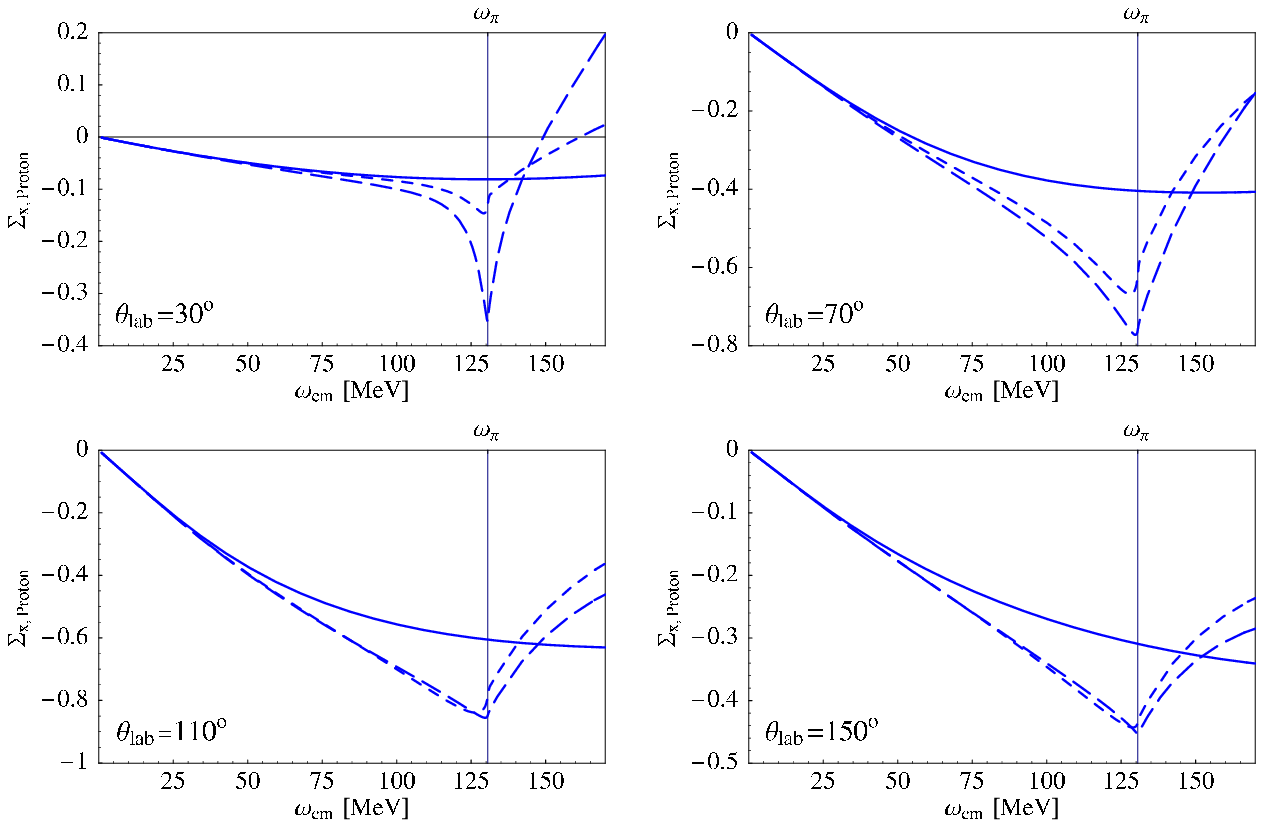} \hq\hq
  \includegraphics[width=0.4\linewidth,bb=110 494 286
  611,clip=]{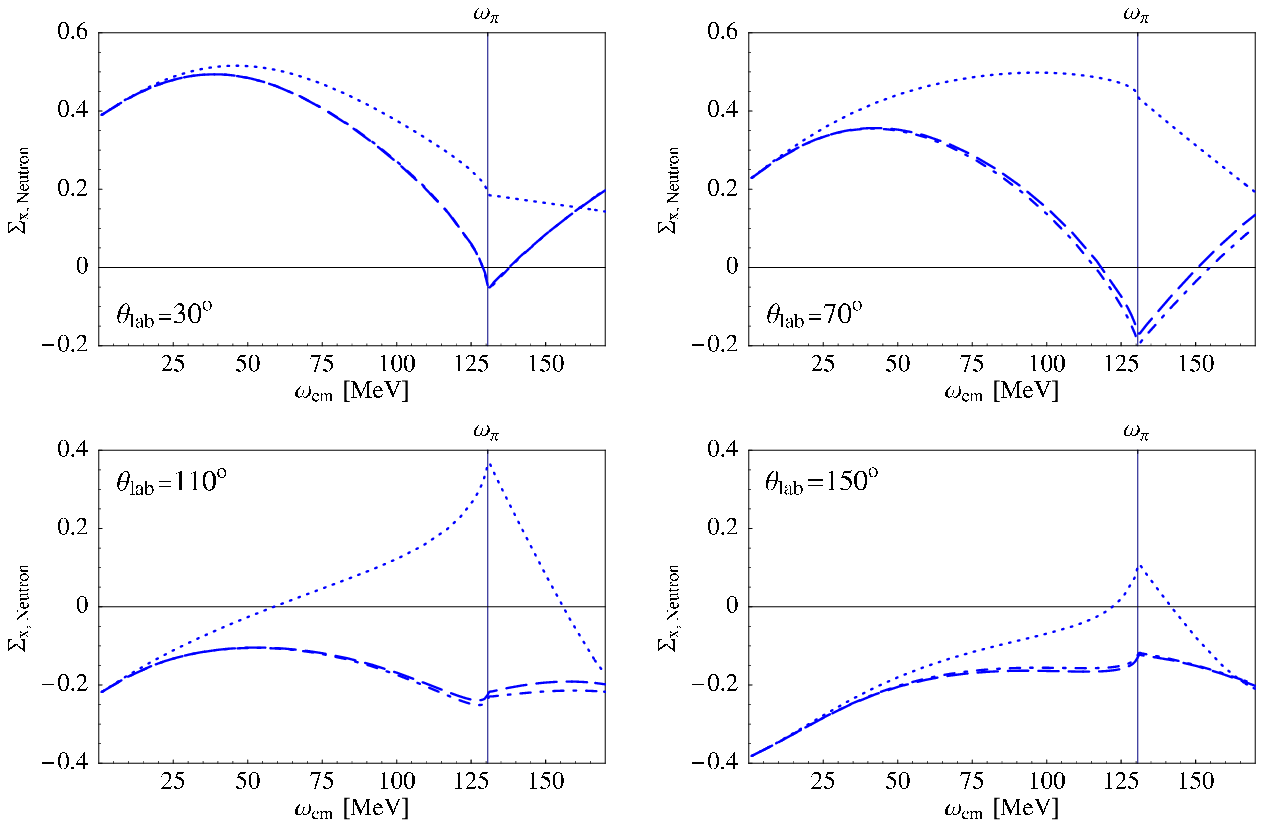}\\[1ex]
  \parbox{14cm} {\footnotesize Fig.~4: Typical sensitivity of the proton
    (left) and neutron (right) Compton scattering asymmetry $\Sigma_x$ on
    $\Delta$ Physics and the spin- and higher polarisabilities,~\cite{polas3}.
    Left, solid lines: no polarisabilities; short dashed: no $\Delta$ Physics;
    long dashed: full $\chi$EFT. Right, dashed lines: full $\chi$EFT result;
    dotted: without spin-polarisabilities; dot-dashed: without quadrupole
    polarisabilities.}
\end{center}

An analysis of Compton scattering via a multipole decomposition at fixed
energies can thus substantially further our knowledge on the spin-structure of
the proton. It will also provide better data on the neutron polarisabilities,
which are known much less accurately than the proton ones.  Double-polarised,
high-accuracy experiments provide thus a new avenue to extract the
energy-dependence of the six polarisabilities per nucleon, both
spin-independent~\cite{polas2} and spin-dependent~\cite{polas3}. A (certainly
incomplete) list of planned or approved experiments at photon energies below
$300\MeV$ shows the concerted effort in this field: polarised photons on
polarised deuterons and ${}^3$He at TUNL/HI$\gamma$S; tagged protons at
S-DALINAC; polarised photons on polarised protons at MAMI; and deuteron
targets at MAXlab.

For example, the asymmetry $\Sigma_x$ between the neutron-spin in positive or
negative $x$-direction in Compton scattering of a circularly polarised photon
with momentum in the $z$-direction shows strong sensitivity on the
spin-polarisa\-bilities $\gamma_i(\omega)$ and on $\Delta$-Physics, while
higher polarisabilities are negligible; see Fig.~4 and~\cite{polas3}. Similar
findings hold for other asymmetries.

\section{Concluding Words}

Dynamical polarisabilities are a concept complementary to \emph{generalised}
polarisabilities of the nucleon, and more directly accessible. The latter
probe the nucleon in virtual Compton scattering, i.e.~with an incoming photon
of non-zero virtuality, and provide information about the spatial distribution
of charges and magnetism inside the nucleon.  \emph{Dynamical
  polarisabilities} on the other hand test the global response of the internal
nucleonic degrees of freedom to a \emph{real} photon of \emph{non-zero} energy
and answer the question, \emph{which} internal degrees of freedom govern the
structure of the nucleon at low energies. They do not contain more or less
information than the corresponding Compton scattering amplitudes, but the
facts are more readily accessible and easier to interpret. Enlightening
insight into the electro-magnetic structure of the nucleon has already been
gained from them, and a host of experimental activities is going to add to
them in the next years.
Last not least, I thank R.P.~Hildebrandt, T.R.~Hemmert and B.~Pasquini for a
fun collaboration!

\end{document}